\documentclass{article}

\usepackage[preprint]{nips_2018}

\usepackage[utf8]{inputenc} 
\usepackage[T1]{fontenc}    
\usepackage{hyperref}       
\usepackage{url}            
\usepackage{booktabs}       
\usepackage{amsfonts}       
\usepackage{nicefrac}       
\usepackage{microtype}      
\usepackage{dirtytalk}
\usepackage[pdftex]{graphicx} 

\title{Dataset Bias in the Natural Sciences: A Case Study in Chemical Reaction Prediction and Synthesis Design}

\author{
  Ryan-Rhys~Griffiths \\
  Department of Physics\\
  University of Cambridge\\
  \texttt{rrg27@cam.ac.uk} \\
\And
Philippe Schwaller \\
Department of Physics \\
University of Cambridge \& \\
IBM Research Zurich \\
\texttt{PHS@zurich.ibm.com} \\
\And
Alpha A. Lee \\
Department of Physics \\
University of Cambridge \\
\texttt{aal44@cam.ac.uk} \\
}

\begin{document}

\maketitle

\begin{abstract}
Datasets in the Natural Sciences are often curated with the goal of aiding scientific understanding and hence may not always be in a form that facilitates the application of machine learning. In this paper, we identify three trends within the fields of chemical reaction prediction and synthesis design that require a change in direction. First, the manner in which reaction datasets are split into reactants and reagents encourages testing models in an unrealistically generous manner. Second, we highlight the prevalence of mislabelled data, and suggest that the focus should be on outlier removal rather than data fitting only. Lastly, we discuss the problem of reagent prediction, in addition to reactant prediction, in order to solve the full synthesis design problem, highlighting the mismatch between what machine learning solves and what a lab chemist would need. Our critiques are also relevant to the burgeoning field of using machine learning to accelerate progress in experimental Natural Sciences, where datasets are often split in a biased way, are highly noisy, and contextual variables that are not evident from the data strongly influence the outcome of experiments.\end{abstract}


\section{Introduction}

Inventing new molecules through synthesis design is a central challenge for chemistry. Computers can process vast numbers of experimental reports and are able to accurately calculate the relative rates of competing reactions. It should therefore be the case that computers have the potential to produce more reliable synthetic routes than humans \cite{2018_Book}. The synthesis design problem as well as the associated problems of reaction prediction and reaction planning are illustrated in \autoref{fig: 1}. The application of Machine learning to this problem has a history at NIPS \cite{2011_Kayala} and has recently been demonstrated to be the state-of-the-art approach both in reaction prediction \cite{2018_Coley_Graph_Conv, 2018_Bradshaw, 2017_Jin} and in synthesis design for reactants \cite{2017_Pande, 2017_Coley_Retro, 2018_Segler}. 

When married together, solutions to the reaction planning, reaction prediction and synthesis planning problems may be used to automatically propose routes to new molecules. Issues in dataset bias however are preventing the attainment of superhuman performance. In this paper, we identify three trends in the application of machine learning with respect to the design of synthetic pathways that may benefit from a change in direction. In section 2 we discuss reagent labelling in the reaction prediction problem and suggest an approach that may bring machine learning systems in line with industrial expectations. In section 3 we discuss the need for outlier detection in noisy reaction prediction datasets whilst in section 4 we highlight the fact that the prediction of reagents in addition to reactants in synthesis design is a key component in the design of a full synthetic route. We conclude by highlighting other areas in the Natural Sciences where dataset bias has been found to stymie progress, emphasizing the generality of the problem.

\begin{figure*}
\centering
{\label{fig: 1}\includegraphics[width=0.75\textwidth]{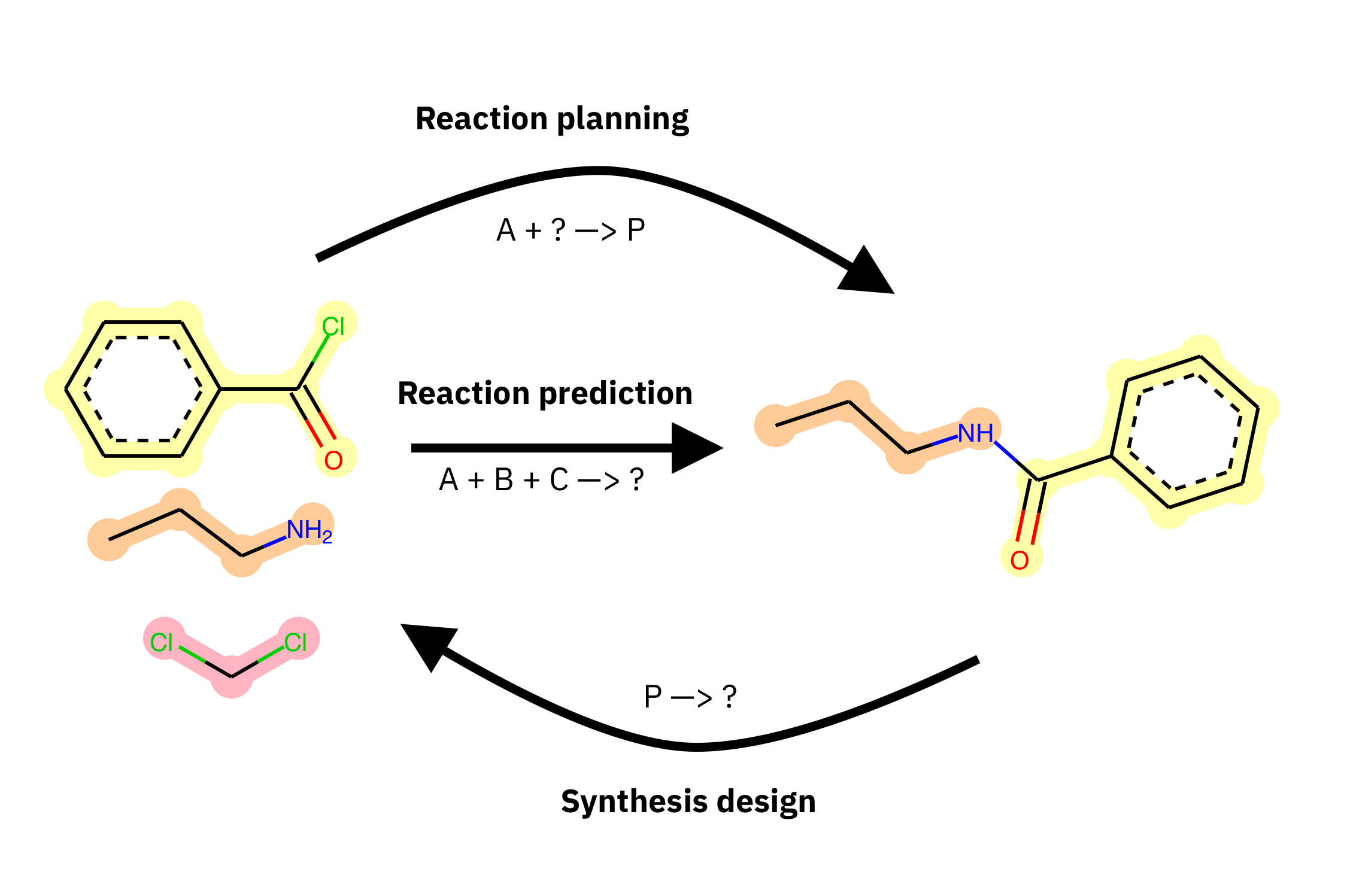}}
\caption{Designing Synthetic Pathways. The reaction planning problem involves finding a set of reagents to transform a given starting material into a given product. The reaction prediction problem concerns predicting the product given a set of reactants and the synthesis design problem involves working backwards from the product towards a set of reactants and reagents.}
\label{fig: 1}
\end{figure*} 

\section{Reagent Labelling in Chemical Reaction Prediction}



According to the IUPAC Compendium of Chemical Terminology a reagent is defined as \say{a test substance that is added to a system in order to bring about a reaction or to see whether a reaction occurs} \cite{Goldbook}. In the chemical reaction prediction literature, the definition of a reagent is more precisely defined to be a compound that does not contribute atoms to the product. 


For a synthetic organic chemist predicting the product of a new reaction, there is no way of knowing a priori which compounds in the reaction mixture will contribute atoms to the product because this is precisely the reaction prediction problem. Therefore, information about which chemicals are reactants and which chemicals are reagents implies prior knowledge about the product (see, e.g. \cite{2016_Landrum}, for algorithms that split reactants and reagents). As such, although the training set can be split into reactants and reagents, splitting the input of the test set into reactants and reagents makes the reaction prediction problem circular because the split can only be done with the answer known a priori. This dataset split is, however, routinely performed in current approaches \cite{2017_Schwaller, 2017_Jin, 2018_Bradshaw, 2019_Anonymous}, where experiments are reported in which reagent labels have been provided to the model at test time. The distinction between providing and not providing reagent labels is illustrated in \autoref{fig: 2}.


\cite{2017_Jin} report results where the top-1 accuracy for the prediction model increases by 6\% when reagents are labelled in the reaction prediction step. \cite{2017_Schwaller} explicitly label reagents using separate tokens. \cite{2018_Bradshaw} input reagent information as a context vector to their model and \cite{2019_Anonymous} report results where improved performance is obtained with labelled reagents. Improved performance is not surprising in this case given that the space of possible products is narrowed through the exclusion of side reactions with the reagent. Since reagent labels are never available before a reaction is carried out, our recommendation would be for machine learning models to be benchmarked exclusively without reagent labelling. This has been done in \cite{2018_Transformer} as well as the human comparison experiments of \cite{2018_Coley_Graph_Conv}.

\begin{figure*}
\centering
{\label{fig: 2}\includegraphics[width=0.5\textwidth]{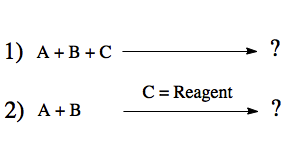}}
\caption{Reagent Labelling: 1) represents the realistic situation where no advance knowledge is available about which of the materials A, B or C acts as reactant/reagent. The question mark represents the product being predicted. 2) represents the output of many recent reaction prediction systems where C has been explicitly labelled as the reagent and is known to contribute no atoms to the product. In this instance, the system may be seen to be receiving some part of the ground truth label for the test-time prediction.}
\label{fig: 2}
\end{figure*} 

\section{Noise in Chemical Reaction Data}


Another difference between domains where machine learning has achieved significant progress, such as image recognition, and chemistry is the prevalence of label noise in chemical data. Image data benefits from negligible label noise, e.g. most images in MNIST and CIFAR-10 are correctly labelled (leading to the recent observation that algorithms which perfectly fit the training set can also achieve good generalisation error \cite{2018_Minorm}). However, datasets in chemistry are often highly noisy. In chemical reaction prediction datasets such as the USPTO dataset of \cite{2012_Lowe}, there are numerous examples in which atoms are unduly present in the product, violating conservation of mass, as illustrated in \autoref{fig: 3}. 


Although the community has, thus far, focused on improving model accuracy, the presence of a significant amount of noise in the dataset suggests that dataset cleaning and outlier removal is perhaps the key stumbling block to achieving superhuman performance. For example, \say{arrow pushing} -- mapping out electron paths -- is a sanity check tool in organic chemistry. The innovative preprocessing step of \cite{2018_Bradshaw} models arrow pushing and prunes the data to consider only reactions that can be explained by linear electron topology. We argue that more effort should go into identifying and removing \say{impossible} chemical reactions from the dataset. We also speculate that a Bayesian approach which models aleatoric uncertainty -- the inhomogeneous distribution of noise in the data -- is an appropriate model to explore \cite{2017_Kendall}. Specifically, the noise can come from (1) incorrect reporting of the structure in patents, (2) errors in transcribing the data into a digital format, as chemical structures extracted from patents are digitised using OCR technologies \cite{2016_Schneider}. For both sources of noise, we argue that a reasonable conjecture is that the noise increases as a function of the chemical complexity of the reactants and reagents and is hence heteroscedastic.

\begin{figure*}
\centering
{\label{fig: 3}\includegraphics[width=\textwidth]{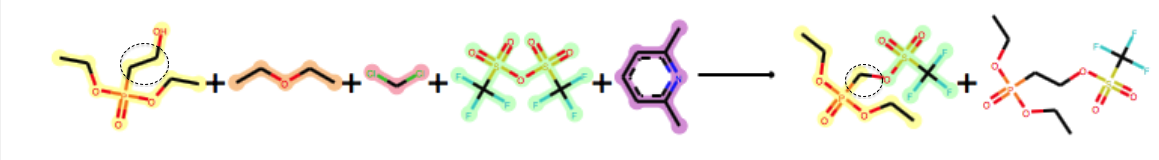}}
\caption{An example in the USPTO dataset of granted patent reactions where conservation of mass is violated. On comparing the circled region in the moiety on the extreme left-hand side of the figure with the moiety immediately to the right of the reaction arrow, we see that an extra carbon atom is present in the reactant that is not present in the product.}
\label{fig: 3}
\end{figure*} 

\section{Achieving the Automation of Synthesis Design}

Although much progress has been made in the prediction of reactants given products \cite{2017_Pande, 2017_Coley_Retro, 2018_Segler}, in order for the prediction to be actionable in the lab, one must predict both reactants and reagents given the products. This is because all chemical molecules can function as a reagent or a reactant depending on the chemical context and separating the two is an artificial construct as discussed above. The distinction between the problem solved by recent approaches and the full synthesis design problem is illustrated in \autoref{fig: 4}. More worryingly, the problem of reagent prediction that is seldom considered by the machine learning community is actually a challenging one -- in fact, many Nobel prize-winning chemical innovation has centred on the discovery of new reagents \cite{Wittig, Suzuki, Sharpless}. As such, there is a gap between machine learning solutions and domain requirements. 



\begin{figure*}
\centering
{\label{fig: 4}\includegraphics[width=0.5\textwidth]{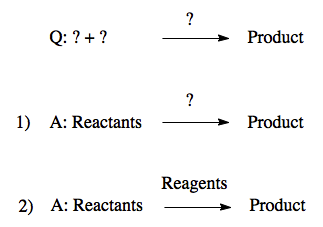}}
\caption{Towards Synthesis Design. The question is how to predict the reactants and reagents given the product. 1) Shows the problem currently being tackled, namely to only predict reactants given the product. The true goal however is 2) Predicting both reactants and reagents given the product in order to enable full automation of synthetic design.}
\label{fig: 4}
\end{figure*} 

\section{Discussion}
 
In this paper, we have focussed on dataset bias in chemical reaction prediction. Dataset bias however, is also prevalent in other areas; in ligand-based drug discovery, it has recently been shown that redundancy in the training and test sets can yield performance measures that do not accurately reflect industrial usage \cite{2018_Atomwise} whilst in the field of solubility prediction, literature values for diclofenac may differ by a factor of 100 \cite{2007_Goodman} and so outlier detection should be an important preprocessing step.

In the field of molecule generation, the gap between machine learning system outputs and industrial requirements may benefit from the presence of more realistic benchmark objective functions. For example, the widely-used penalised logP objective \cite{2018_Design, 2017_Grammar, 2017_Griffiths, 2018_Molgan, 2018_Jin, 2018_Kajino, 2018_Zhou, 2018_You, 2017_Yang, 2018_Popova, 2018_Entangled} is likely to be too smooth a function of the molecular representation to resemble chemically interesting objective functions. 

In accordance with the No Free Lunch Theorems \cite{1994_Schaffer, 1997_Wolpert, 2001_Wolpert}, domain knowledge about the properties of industrially relevant objective functions may yield better heuristics for algorithm design. Objectives such as IC50 or binding affinity \cite{2018_Harel, 2018_Aumentado, 2018_Blaschke} are likely to be more interesting for drug discovery. To score highly for these properties, a certain level of specificity is required of the molecule and so one would expect the objective function to have sharp minima as opposed to broad minima as in the case of logP where similar molecules are expected to exhibit similar values of logP. Indeed positive work on correcting this trend of logP optimisation benchmarking has already been suggested by \cite{2018_Neil}, who recommend a series of industrially relevant objectives.

\newpage

\section{Conclusion}

In this paper we have highlighted three trends within the fields of chemical reaction prediction and synthesis design related to dataset presentation that need to be rethought: reagent labelling, outlier detection and reagent prediction. Our recommendations should hopefully redirect focus to problems that are of industrial relevance. We also discussed how problems associated with dataset presentation extend beyond the domain of reaction prediction. Perhaps outside the scope of Machine Learning but important to reaction prediction, we note that the chemistry literature is biased towards successful reactions and hence another dataset bias is the absence of data on reactions where the reactants do not react. Without non-reactions in the dataset, algorithms are biased towards predicting a chemical change in the reactants, whereas in reality it is not true that randomly mixing chemicals will always cause a chemical transformation. We suggest that a way to tackle this is through encouraging the adoption and sharing of Electronic Lab Notebooks in academic synthetic organic chemistry.




\small

\bibliography{References/example_paper}

\end{document}